\documentclass[ onecolumn, superscriptaddress, floatfix, aps ]{revtex4-2}
\usepackage[utf8]{inputenc}
\usepackage[letterpaper,top=2cm,bottom=2cm,left=3cm,right=3cm,marginparwidth=1.75cm]{geometry}

\usepackage{amssymb}
\usepackage{amsmath}
\usepackage{subcaption}
\usepackage{graphicx}
\usepackage[capitalize]{cleveref}
\usepackage{gensymb}\usepackage{float}

\begin{document}

\title{Dispersion Engineered Metastructures Enabling Broadband Angular Selectivity}
\author{Phillippe Pearson}
\affiliation{Thomas J. Watson Laboratories of Applied Physics, California Institute of Technology, Pasadena, California 91125 USA}

\author{Zhaowei Dai}
\affiliation{Department of Electrical and Computer Engineering, Yale University, New Haven, CT 06511 USA}
\affiliation{Energy Sciences Institute, Yale University, New Haven, CT 06511 USA}
\author{Yiran Gu}
\affiliation{Thomas J. Watson Laboratories of Applied Physics, California Institute of Technology, Pasadena, California 91125 USA}
\author{Owen D. Miller}
\affiliation{Department of Electrical and Computer Engineering, Yale University, New Haven, CT 06511 USA}
\affiliation{Energy Sciences Institute, Yale University, New Haven, CT 06511 USA}
\affiliation{Department of Applied Physics, Yale University, New Haven, CT 06511 USA}
\author{Andrei Faraon}
\affiliation{Thomas J. Watson Laboratories of Applied Physics, California Institute of Technology, Pasadena, California 91125 USA}

\begin{abstract}

Angle-selective optical devices are of importance to several applications such as photovoltaics, high-sensitivity photodetectors and displays. There are several approaches to realizing angular selectivity, but it remains challenging to obtain isotropic responses over large spectral bandwidths in optically thin structures. We introduce a dispersion engineering approach coupled with topology optimization to design 2D metastructures, leveraging guided-mode resonances (GMRs), that exhibit isotropic angular selectivity over relative bandwidths of approximately 20\%. We experimentally demonstrate metastructures with complementary angular selectivities, either scattering light strongly near normal incidence and transmitting efficiently at higher incident angles, or vice versa. A key finding is that these designs enable operation over spectral bandwidths greater than the GMR linewidths would suggest, a result of carefully tailored interactions between the Fabry-Perot background and resonantly scattered light. This work marks a significant step forward for the realization of broadband, angle-selective scattering in readily fabricated structures of subwavelength thickness, and enables new possibilities in sensing, analog information processing, high-efficiency photovoltaics, and displays.

\end{abstract}

\maketitle

\section{Introduction}

Filtering light based on its propagation direction is fundamental for many applications and has been explored extensively over the past several decades. There have been many proposed and experimentally realized implementations of angle-selective optical devices based on, for instance, multi-layer thin film stacks, generalized Brewster angles, epsilon-near-zero (ENZ) materials, volume Bragg gratings (VBGs), topological transitions, photonic crystals, and guided mode resonances (GMRs) \cite{shen_optical_2014,g_w_mbise_angular_1997,wesemann_selective_2019, cotrufo_dispersion_2023, qu_polarization-independent_2018,xu_broadband_2021,kwon_nonlocal_2018,kogelnik_coupled_1969,huo_angular_2018, hamam_angular_2011, xiong_augmented_2021}, each presenting their own set of drawbacks and benefits. However, a fundamental challenge underlying most of them is the difficulty in decoupling the correlation between the angular and spectral bandwidths, a natural consequence of momentum conservation in periodic structures \cite{jacob_design_2000}. A well-known example is Kogelnik's coupled-wave theory for VBGs that prescribes relationships between the angular and spectral ranges over which efficient Bragg diffraction can occur \cite{kogelnik_coupled_1969}. Furthermore, devices demonstrating broadband angular selectivity are typically very thick, often requiring either many tens of thin film layers or high aspect ratio patterning. Thus, there are a lack of suitable approaches for designing structures of subwavelength thickness with uniform angle-selectivity over wide spectral bands.

Guided-mode resonances in high-contrast dielectric gratings have been studied in detail and have demonstrated capabilities as narrow- and broadband spectral filters \cite{magnusson_physical_2008,c_f_r_mateus_broad-band_2004, d_rosenblatt_resonant_1997,shokooh-saremi_wideband_2008,wang_theory_1993, fan_analysis_2002}. Nevertheless, their potential for broadband angle-selectivity has seldom been explored, and most work has sought to eliminate any variation in scattering as a function of incident angle \cite{ko_properties_2016,boonruang_broadening_2007}. Inspired by the strong nonlocality and spectral isolation of low-order guided slab modes folded above the light line (i.e. GMRs), we present a dispersion engineering strategy coupled with topology optimization to realize metastructures exhibiting broadband, isotropic angular selectivity. We show that these structures exceed the bandwidth limitations of certain VBGs while scattering isotropically and being an order of magnitude thinner. This paper begins by introducing the theoretical framework underpinning the functionality of the devices to be presented and demonstrated. Specifically, temporal coupled-mode theory (CMT) elucidates necessary conditions on the symmetry of the GMRs and background scattering to achieve angular selectivity. We apply these ideas to the design of a simple 1D dielectric grating that suppresses transmission near normal incidence and transmits light at larger angles of incidence (\cref{fig:fig1}b left panel). Then, we introduce topology optimization as a means to extend the functionality to 2D metastructures with isotropic scattering response. Finally, we turn our attention to an angular response that transmits near normal incidence while scattering strongly away from it, as depicted in the right panel of \cref{fig:fig1}b -- complementary functionality that may find applications in augmented reality systems, photovoltaics, and photodetectors with enhanced signal-to-noise ratios.

\section{Results}

\subsection{Design principles for 1D grating structures}

\begin{figure}
    \centering
    \includegraphics[width=0.6\linewidth]{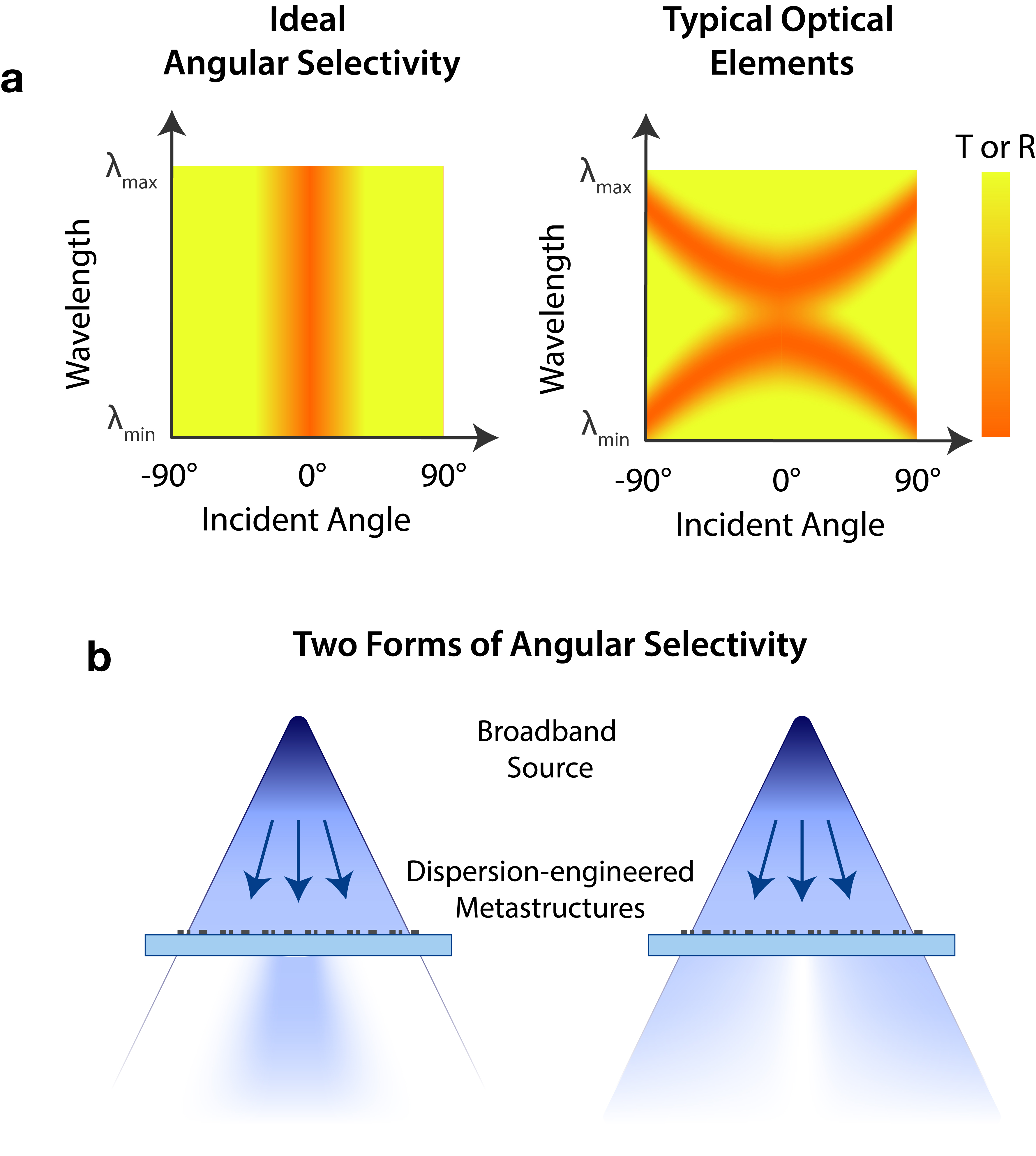}
    \caption{\textbf{a)} The left panel shows an ideal angle-selective response while the right one represents what one typically encounters in practice. The color scale can indicate either transmittance or reflectance. \textbf{b)} The two kinds of angular selectivity under consideration. On the left is a situation where light exhibits specular transmission near normal incidence and strong scattering at higher angles. On the right, the opposite case is shown.}
    \label{fig:fig1}
\end{figure}

We begin by presenting an approach for scattering light strongly within some angular range near normal incidence, while transmitting specularly with high efficiency outside of it. This idea is summarized in \cref{fig:fig1}a, which shows an idealized broadband angle dependent scattering response. Clearly, this is very difficult to realize in practice over an appreciable spectral bandwidth, and ultimately in most resonant structures the dispersion relation more closely resembles the right panel of \cref{fig:fig1}a. A wavelength-independent scattering resonance implies that the resonance responsible for it runs parallel to the light line defined by $\omega=c k_x$, where $c$ is the speed of light in vacuum, $\omega$ is the angular frequency and $k_x$ is the in-plane wave vector. This motivates us to search for resonances that get as close as possible to the slope of the light line over a broad spectral band. For angles near normal, this requires large group velocities. Ideally, this resonance band should also be spectrally isolated such that the resulting scattering band is unperturbed.

Guided mode resonances (GMRs) originate from guided slab bands that are periodically perturbed such that they couple to free space light with a finite quality factor $Q$. Stated differently, the periodic modulation folds the guided band above the light line. In their simplest implementation, GMRs are found in one dimensional (1D) dielectric grating structures on a substrate and can be designed to host highly dispersive resonances with a wide range of quality factors \cite{wang_theory_1993}. Guided slab bands of the lowest order are also the most spectrally isolated, and as such, we focus on them to realize angle selective scattering. Near $k_x=0$ the folded band splits into an upper and lower branch \cite{fan_analysis_2002,wang_theory_1993}, resulting in two closely spaced resonances whose scattering response we model with a CMT to capture the symmetries of the resonances \cite{wonjoo_suh_temporal_2004}. Considering a dielectric grating structure surrounded by vacuum such that it possesses reflection symmetry plane ($\sigma_y$) bisecting its thickness at $y=0$, resonances may be classified as odd or even with respect to this plane \cite{fan_analysis_2002}. Since the GMR branches originate from the same guided band, their field profiles must have identical symmetry with respect to $\sigma_y$. There is no such restriction on their $\sigma_x$ symmetries, as we will see below. Thus, these are non-orthogonal modes, and their transmission characteristics are captured by the following expression,
\begin{equation}
t = t_d \mp \frac{(r_d \pm t_d)(\gamma_1 j(\omega-\omega_2) + \gamma_2 j (\omega-\omega_1))}
{(j(\omega-\omega_1) + \gamma_1)(j(\omega-\omega_2) + \gamma_2) - \gamma_1 \gamma_2},
\label{cmt_t}
\end{equation}
where $\omega_{1,2}$ are the resonant frequencies and $\gamma_{1,2}$ are the decay rates, with their dependence on $k_x$ omitted for notational simplicity. The background scattering processes through the slab are transmission, with coefficient $t_d$, and reflection, with coefficient $r_d$. The upper and lower signs are used, respectively, for modes with even and odd mirror symmetry. As a simple model, we assign linear dispersion relations of the form $\omega_r=\omega_0 \pm v_g k_x$ to the upper (+ sign) and lower (- sign) bands and set them to be degenerate at $k_x=0$ with identical decay rates. For $r_d$ and $t_d$, we use the analytical expressions for the reflected and transmitted fields for a Fabry-Perot (FP) resonator parameterized by the thickness and effective refractive index of the patterned slab \cite{hecht_optics_2002}. 

\begin{figure}
    \centering
    \includegraphics[width=\linewidth]{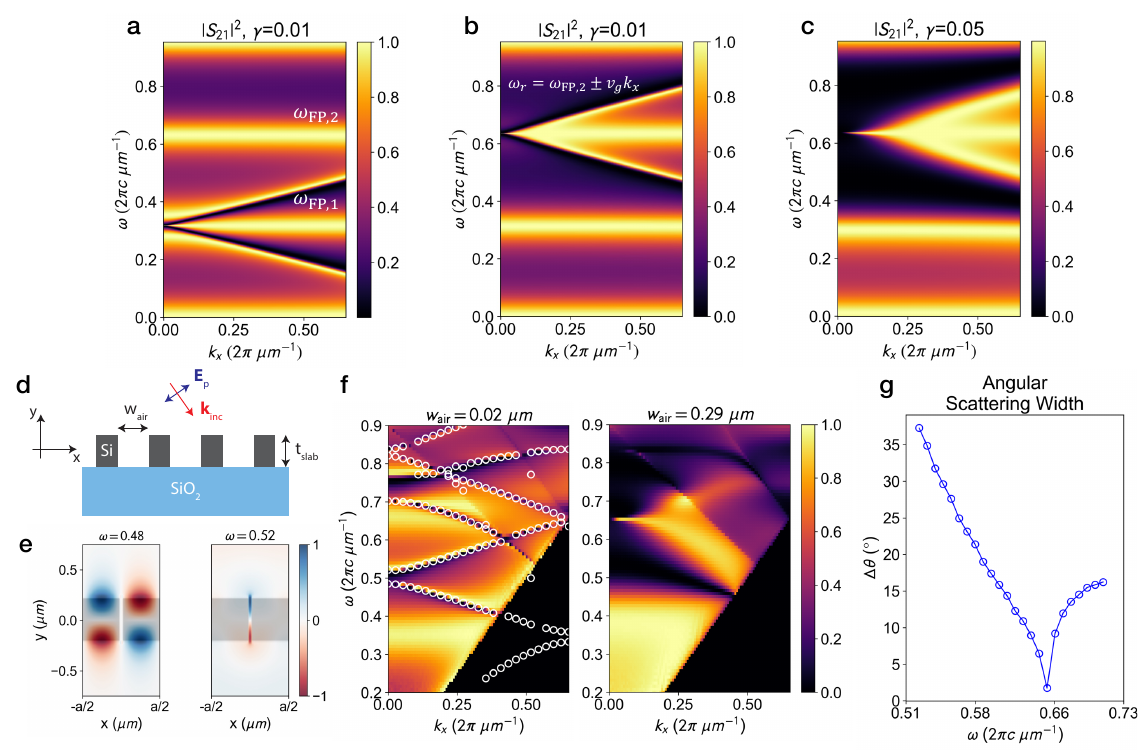}
    \caption{\textbf{Dispersion engineering in 1D dielectric gratings. a)} Transmittance predicted by coupled-mode theory for two non-orthogonal modes of odd symmetry coupled to a Fabry-Perot background. The $\omega_{\mathrm{FP}}$ labels represent the first two FP modes of the slab, which is parameterized by a thickness of 0.45 $\mu m$ and an effective index of 3.5, similar to silicon. The resonances are both tuned to $\omega_{\mathrm{FP,1}}$ when $k_x=0$ The predicted transmittance of the same modes but instead tuned to $\omega_{\mathrm{FP,2}}$ for \textbf{b)} $Q_{\mathrm{mode}}=31.7$ and \textbf{c)} $Q_{\mathrm{mode}}=6.3$. \textbf{d)} Cross section of the extruded grating geometry consisting of patterned Si on a glass substrate illuminated with a plane wave. \textbf{e)} Simulated $E_x$ profiles of the upper and lower branches of the lowest order GMR band at $k_x=0$ in a structure with $w_{\mathrm{air}}=0.02$ $\mu m$. The field amplitudes are normalized to their respective largest absolute value. \textbf{f)} Simulated transmittance with FDTD of the grating with $w_{\mathrm{air}}=0.02$ $\mu m$ (left) and $w_{\mathrm{air}}=0.29$ $\mu m$ (right). In the left panel, the resonant modes of the structure are shown as white circles overlaid on the transmittance; they are calculated by analyzing the decaying tail of the field amplitudes after exciting the grating with a point source. \textbf{g)} The angular width over which transmittance is strongly suppressed, defined as the angle where $T(k_x)=T_{\mathrm{max}}/2$ for each $\omega$ in the spectral band of interest. The average value is 18.1\degree.}  
    \label{fig:fig2}
\end{figure}

\Cref{fig:fig2}a shows $T(\omega,k_x)$ calculated from \cref{cmt_t} where odd symmetry GMRs with $\gamma = 0.01$ are tuned exactly to the first FP mode $\omega_{\mathrm{FP,1}}$ at $k_x=0$ resulting in strong reflectance and angular selectivity within this bandwidth. However, when the GMRs are instead tuned to $\omega_{\mathrm{FP,2}}$ we observe strong reflectance that spans a spectral band much wider than would be expected based on the $Q$ of the individual GMRs (\cref{fig:fig2}b). The striking contrast in the reflectance band is attributed to the fact that light transmitted on resonance by adjacent FP modes is $\pi$ out-of-phase. A more detailed discussion of this effect is included in the Supplement. Now, \cref{fig:fig2}c demonstrates that increasing $\gamma$ (i.e. reduction of $Q$) further suppresses transmission and induces angular selectivity due to the asymmetry of the Fano lineshape on either side of $\omega_{\mathrm{FP,2}}$. The idea of using GMRs as broadband, high-reflectivity mirrors has been studied \cite{heo_tailoring_2019,shokooh-saremi_wideband_2008}, yet their application to angular selectivity has not been explored extensively. The CMT calculations thus suggest that an appropriate alignment of the GMR bands and FP peaks can produce a broadband angle-selective response.

We use the insights gained from this analysis to design a 1D dielectric grating with broadband angular selectivity whose geometry is shown in \cref{fig:fig2}d. The structure consists of a silicon (Si) grating with a thickness of $t_{\mathrm{slab}}$=0.45 $\mu m$ and period of $d$=0.76 $\mu m$ on a glass (SiO$_2$) substrate. In the absence of a substrate, the lowest frequency transverse magnetic (TM or $p$-polarized) mode of this structure has odd mirror symmetry with respect to the $y$=0 plane. Despite the addition of a substrate, the reflection symmetry is only slightly perturbed, and the design strategy outlined above remains valid. The $p$-polarized transmittance of the structure with $w_{\mathrm{air}}$=0.02 $\mu m$ is simulated with the finite difference time domain method (FDTD) as implemented in the open-source package MEEP and is presented in \cref{fig:fig2}f along with its overlaid resonant modes \cite{oskooi_meep_2010}. These modes are calculated by exciting the structure with an $H_z$-polarized dipole and analyzing the temporal decay of the fields with a filter diagonalization algorithm \cite{fan_analysis_2002}. We can see that the lowest order band is folded at the edge of the first Brioullin zone and becomes a GMR above the light line, intersecting with $k_x$=0 near $\omega$=0.5 and splitting into an upper and lower branch. The $E_x$ field profiles of these resonances are shown in \cref{fig:fig2}e and clearly demonstrate their odd symmetry. As expected, the field of the lower and higher frequency modes is mostly confined to the Si and air regions, respectively, which introduces different $\sigma_x$ symmetry for each. Consequently, the lower branch becomes a symmetry-protected bound state in the continuum at normal incidence. We also observe the first and second FP peaks in the simulated transmittance. To tune the odd GMR band near the second FP peak and decrease its $Q$, we widen the etched air trench to $w_{\mathrm{air}}$=0.29 $\mu m$. The transmittance of this structure is shown in the right panel of \cref{fig:fig2}f and, as suggested by CMT, we observe broadband transmittance suppression with a strong angular dependence. Specifically, over a fractional bandwidth of approximately $\Delta\omega=25\%$, the average full-width half maximum (FWHM) of the \textit{angular} rejection band (using $k_x=\omega\sin\theta$) is approximately +/- 18\degree. This is visualized more explicitly in \cref{fig:fig2}g, which shows the FWHM angular rejection widths computed within the spectral band of interest in \cref{fig:fig2}f. The experimentally measured transmittance of the fabricated structure can be found in Fig. S5 and is in good agreement with the simulation. In the Supplement, we present an analysis comparing this result to the well-known relationship between angular and spectral scattering bandwidths of reflective volume Bragg gratings (VBGs) and find that, given some spectral width, GMR-based structures can provide narrower angular selectivity in certain cases while being dramatically thinner than typical VBGs.

\subsection{Topology optimization for 2D isotropic selectivity}
\begin{figure}[]
    \centering
    \includegraphics[width=0.85\linewidth]{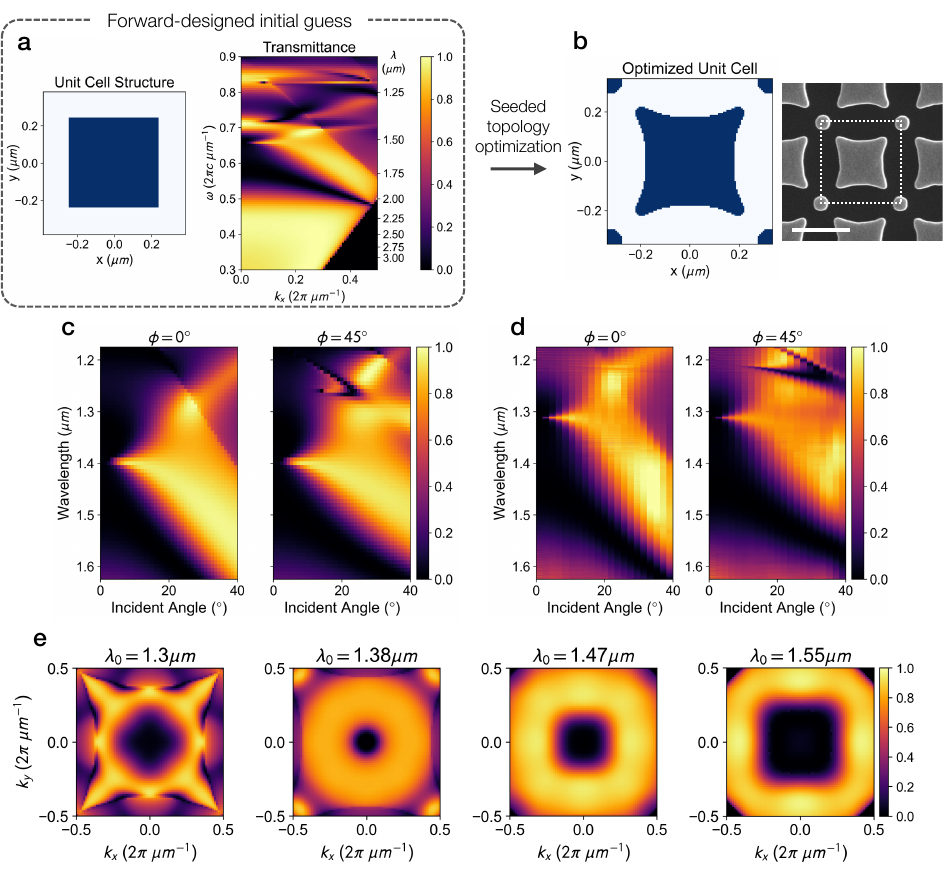}
    \caption{\textbf{Isotropic angular selectivity with topology optimization. a)} Forward designed 2D grating structure obtained through a ``naive'' extension of the 1D version. On the left, a bird's eye view of the unit cell is shown with a period of 0.76 $\mu m$ and $w_{\mathrm{air}}=0.29$ $\mu m$. The transmittance calculated with RCWA is shown on the right. \textbf{b)} Topology-optimized unit cell and an SEM of the fabricated metastructure. The thickness and period are 0.475 $\mu m$ and 0.705 $\mu m$, respectively. Simulated \textbf{c)} and experimentally measured \textbf{d)} transmittance of the topology-optimized structure as a function of wavelength and incident polar angle for $\phi=0\degree$ and $\phi=45\degree$. \textbf{e)} Simulated transmittance as a function of $k_x$ and $k_y$ for four wavelengths.}
    \label{fig:fig3}
\end{figure}

Although the structures considered in the previous section produce broadband angular selectivity, this response exists only along the $x$-direction. We use topology optimization to realize isotropic angular selectivity starting with a 2D grating structure that is a direct extension of the structure in \cref{fig:fig2}. A naive guess consists of simply adding a periodic modulation running orthogonal to the grating of the previous section. However, this immediately introduces a challenge due to the folding of two additional bands along the $k_y$ direction that end up close to the GMR bands of interest near $k_x=0$ (Ref.~\cite{sakoda_optical_2005}). These bands also have lower group velocity along the $k_x$ direction such that they intersect the smoothly varying spectrum of the 1D structure. \Cref{fig:fig3}a shows a bird's eye view of the unit cell of the same structure as in \cref{fig:fig2}d but with an additional grating pattern in the orthogonal direction (equivalent to the square of Si in \cref{fig:fig3}a). The transmittance of this structure is shown to its right and we can indeed see the presence of the flat $k_y$ bands that limit its spectral bandwidth. Furthermore, simulations along the $\phi=45\degree$ azimuth reveal a significantly different angular response, highlighting the lack of isotropy (Fig. S6).

To design a 2D structure that maximally recovers the performance of its 1D counterpart, we turn to topology optimization \cite{molesky_inverse_2018,su_inverse_2018,lalau-keraly_adjoint_2013}. This frames the design problem as an optimization where the final structure is obtained such that it minimizes a figure of merit (FoM) capturing the desired optical response. We discretize the unit cell onto a grid and use a differentiable rigorous coupled-wave analysis (RCWA) solver \cite{jin_inverse_2020} to compute the transmittance of the structure such that we can easily obtain the gradients of the FoM with respect to the unit cell permittivity. Based on these gradients, the unit cell permittivity is iteratively adjusted to minimize the FoM. Instead of starting from randomly drawn unit cell permittivities, as is typical in many topology optimization schemes, we use the ``forward design" of \cref{fig:fig3}a as a starting point, leveraging the theoretical framework of the previous section to guide the optimization. The minimum feature size and binarization are enforced at each iteration with a Gaussian blur and projection filter, respectively \cite{wang_projection_2011,li_inverse_2022,wang_design_2022}. The target transmission is set to be $T(\theta) = 1-\exp(\theta/2\sigma_{\theta})$ with $\sigma_{\theta}=10\degree$ and the optimization is carried out for wavelengths spanning 1.3-1.7 $\mu m$, $\phi$=[0\degree,45\degree], and $\theta$ from 0\degree to 25\degree. We have described this method in a prior publication \cite{pearson_inverse-designed_2025} and more details can be found in the Supplement. 

The topology-optimized structure is shown in \cref{fig:fig3}b with the nominal pattern on the left and an SEM image of the fabricated device on the right. The Si is targeted for a thickness of 475 nm with a period of 705 nm, a combination which was found to produce good optimization results. Details of the fabrication process can be found in the Methods section. The simulated transmittance along $\phi$=0\degree and $\phi$=45\degree as a function of wavelength and incident polar angle (\cref{fig:fig3}c) indicates broadband reflection at normal incidence with minimal effects from the additional $k_y$ folded bands present in \cref{fig:fig3}a. The topology optimization procedure produces a metastructure that reduces their effects and has a strong angle-selective response over a fractional bandwidth of nearly 20\%. For the case of $\phi=0\degree$, the angular extent is bounded by the light line of the glass substrate, where at large angles of incidence light is transmitted into higher diffraction orders within the substrate. The experimentally measured angle-resolved transmittance presented in \cref{fig:fig3}d confirms the broadband angular selectivity predicted in simulation. The data is collected by directing light from a supercontinuum laser (SuperK) through the device mounted on an automated rotation stage and then to a grating spectrometer (see Supplement for additional details). Despite an overall blue shift attributed to refractive index variations in the deposited Si, the measured spectral features agree with our simulations. Critically, the transmittance is highly isotropic, which is confirmed by simulations over a 2D grid of $k_x$ and $k_y$ points for four wavelengths (\cref{fig:fig3}e). These results highlight the effectiveness of using topology optimization in conjunction with physics-based design principles, rather than as a purely black box tool.

\subsection{Bandpass angular selectivity}

\begin{figure}
    \centering
    \includegraphics[width=\linewidth]{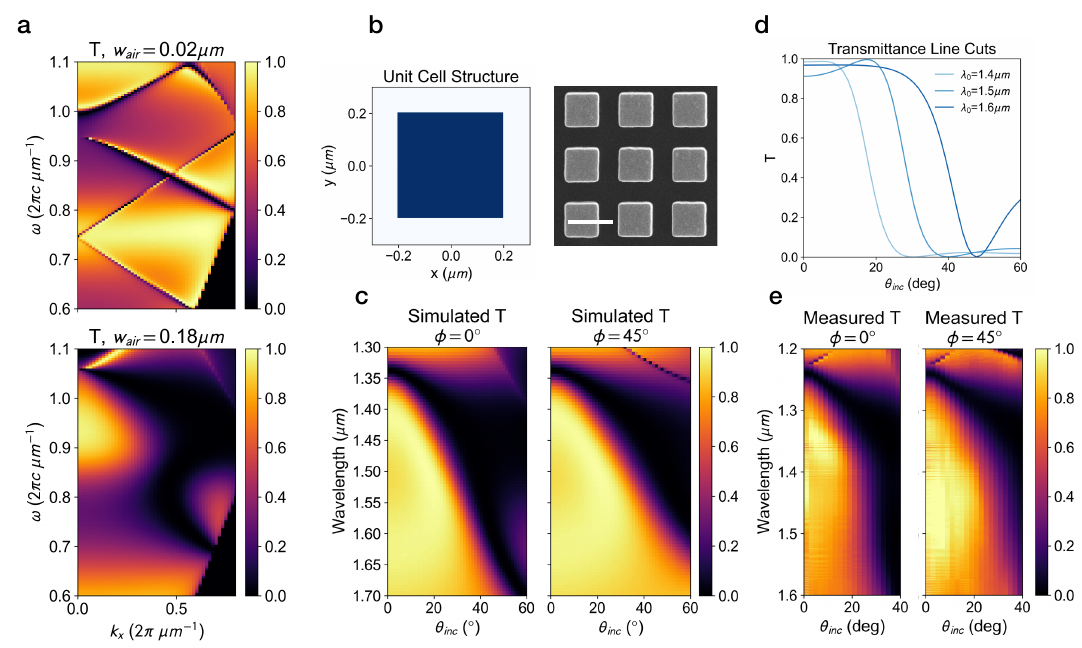}
    \caption{\textbf{Band-pass angular selectivity. a)} Simulated transmittance (RCWA) for a 1D grating structure with a thickness and period of 0.4 $\mu m$ and 0.45 $\mu m$, respectively, for two different values of $w_{air}$. \textbf{b)} Nominal and fabricated unit cell of 2D metastructure. \textbf{c)} Simulated transmittance along $\phi=0\degree$ and $\phi=45\degree$. \textbf{d)} Line cuts along $\theta_{inc}$ of transmittance for three wavelengths. \textbf{e)} Experimentally measured transmittance for the structure in \textbf{b)}.}
    \label{fig:fig4}
\end{figure}

So far, we have explored structures that reflect strongly over some desired angular band and transmit light outside of it. However, several applications may benefit from the complementary functionality where light is transmitted within some angular range and reflected without \cite{fan_broadband_2025,kosten_highly_2013,ulbrich_enhanced_2010,peters_photonic_2009}. An example is the reduction of the rainbow effect in see-through augmented reality headsets where an out-coupling grating is used to direct light towards the eye  \cite{li_switchable_2023,ding_waveguide-based_2023,rolland_waveguide-based_2024}. Ambient light incident at large angles can couple into the eye through a highly dispersive grating order, causing a rainbow-like pattern for the user. A band pass angle-selective optical element designed to reject such ambient light could mitigate this effect. To design an angular transmission band centered at normal incidence, we make a simple modification to the dispersion engineering approach described above. Considering a 1D Si grating on a glass substrate (as in Section A) with a thickness and period of 0.4 $\mu m$ and 0.45 $\mu m$, respectively, \cref{fig:fig4}a shows its transmittance for two different values of $w_{\mathrm{air}}$. In the upper panel, a small perturbation of 20 nm is used and we see that near $k_x=0$ the GMR band is nearly resonant with the second FP peak. When the perturbation is widened to 180 nm, the $Q$ of the GMR decreases and it is blue-shifted relative to the FP peak. In this way, there is large transmission at normal incidence because the lower GMR branch intersects the FP mode at a non-zero $k_x$. This results in the transmittance spectrum shown in the lower panel of \cref{fig:fig4}a that exhibits an angle-selective spectral bandwidth of $\Delta\omega_f\approx$ 15\%.

As before, we extend this scattering behavior to be isotropic by moving to the 2D periodic structure shown in \cref{fig:fig4}b. After scaling and fine tuning the geometry in \cref{fig:fig4}a for good performance near $\lambda_0$=1.5 $\mu m$, the thickness, period, and Si width are 0.7 $\mu m$, 0.6 $\mu m$, and 0.4 $\mu m$. This structure leads to the simulated transmittance of \cref{fig:fig4}c plotted now as function of wavelength and incident angle for two azimuthal directions. The broadband angular selectivity is evident and features a spectral bandwidth nearly identical to the 1D structure of \cref{fig:fig4}a. Furthermore, the response is highly isotropic as demonstrated by the similarity in transmittance spectra for both azimuthal directions shown. \Cref{fig:fig4}d presents three line cuts from the left panel of \cref{fig:fig4}c that highlight the strong contrast between the angular transmission and rejection bands. Finally, the measured angle-resolved transmittance (SEM in \cref{fig:fig4}b) agrees well with the simulation results, aside from a spectral blue shift, as seen in \cref{fig:fig4}e. In this case, the transition from a 1D to 2D structure has minimal effect on the transmission spectrum primarily because the GMR band is detuned from the FP mode at $k_x=0$. This means that the bands folded in from the $k_y$ direction are less likely to appear within the spectral region of interest. Thus, through a simple parameter sweep we achieved a broadband and isotropic angle dependence while forgoing topology optimization.

For comparison, recent work has realized angular selectivity through the generalized Brewster effect in stacks of alternating thin films \cite{shen_optical_2014,qu_polarization-independent_2018}. These structures achieve wider angle-selective spectral bandwidths but are much thicker than our single-layer metastructures. Specifically, for an angular transmission window centered near 55\degree, almost 100 thin film layers are required, and for similar performance in an angular window centered at normal incidence, a structure with more than 1000 layers was proposed. In contrast, our structure is approximately $\lambda_0$/2 in thickness, orders of magnitude more compact.

\section{Discussion and Conclusion}

 To better understand the design tradeoffs inherent to our results, in the Supplement we present a derivation starting from \cref{cmt_t} and arrive at a necessary condition for broadband reflectance when two odd resonances of identical $Q$ are tuned to be degenerate with an odd-numbered FP mode:
\begin{equation}
    \frac{4r\gamma n_{\mathrm{eff}} t_{\mathrm{slab}}}{1-r^2} \geq 1,
    \label{ext_condition}
\end{equation}
where $\gamma$ is the decay rate of the resonance and $r$ is the Fresnel reflectivity of the effective dielectric slab that models the background scattering, characterized by its index $n_{\mathrm{eff}}$ and thickness $t_{\mathrm{slab}}$. A similar result was obtained in \cite{heo_tailoring_2019} when considering a single resonance, but here we extend these ideas to realize broad spectrum, isotropic angular selectivity. This condition highlights several tradeoffs inherent to our design approach. First, it shows that increasing $\gamma$ (i.e. reducing $Q$) can directly lead to broadband reflectance, consistent with our approach of increasing the perturbations in our structures to reduce $Q$. Importantly, this decrease leads to a reflectance band much wider than the $Q$ of the resonance would suggest and is thus fundamentally distinct from the broadening of a typical Lorentzian lineshape. Second, \cref{ext_condition} suggests that increasing $n_{\mathrm{eff}}$ (and thus $r$) also promotes broadband reflectance because the finesse of the FP cavity increases with $r$ and provides stronger reflectance modulation.

These results seemingly suggest that structures comprised of high index dielectric materials are better suited for broadband angular selectivity. However, as $n_{\mathrm{eff}}$ grows, the group velocity of the resonances decreases, which necessarily increases the lower bound on the average angular rejection width (see Supplement) making it more difficult to achieve broadband selectivity over a narrow angular range. As a consequence, there is a fundamental tradeoff between the reflectance bandwidth and the average width of angular rejection or transmission. The above result also implies that designing such a structure to work for visible wavelengths, where high-index and low-loss materials are few and far between, is more challenging. Titania (TiO$_2$) is often used as a transparent dielectric for these wavelengths and has a refractive index near 2.5 \cite{siefke_materials_2016}. In the Supplement, we demonstrate a TiO$_2$-based angle-selective structure designed to operate near $\lambda_0$=500 nm and we find that it is indeed more difficult to achieve broadband reflectance with this material as predicted by \cref{ext_condition}. Nevertheless, this structure produces angular rejection spanning over 100 nm while being exceptionally thin compared to other solutions.

In summary, we have presented a method for designing metastructures with strong angular selectivity operating over wide spectral bands. Our results rely on careful engineering GMR and FP modes whose interactions can suppress transmission near or away from normal incidence. Combining an intuitive theoretical framework with gradient-based topology optimization enables freeform 2D metastructures that can outperform reflective VBGs while also being highly isotropic. This approach is applicable to a wide range of material platforms and wavelength bands. In the future, exploring resonant phenomena with high group velocities is of interest in order to further probe the limits of scattering responses that are spectrally broad and angularly narrow. These may include effects related to surface lattice resonances or low-loss ENZ materials (or metamaterials) \cite{bin-alam_ultra-high-q_2021,hrabar_ultra-broadband_2013,wang_gain-assisted_2000}. We believe that this work may lead to exciting new opportunities in augmented reality, optical information processing, and photovoltaics.

\section{Methods}

The devices presented in this work are fabricated in amorphous Si (a-Si) grown by plasma-enhanced vapor deposition (PECVD) on a glass substrate (500 $\mu m$ thick). The patterning process starts by sonicating the sample in acetone and rinsing with isopropyl alcohol, followed by oxygen plasma. After baking the sample on a hotplate at 180$\degree$C, ZEP520A (Zeon Corporation) is spin-coated on the silicon surface and baked once more. Next, a 15 nm layer of chromium is evaporated on the resist to act as a conductive layer. The pattern is written with an EBPG 5200 (Raith) EBL system. Following exposure, chromium is removed with a wet etching process. The resist is developed in ZED (Zeon Corporation) and rinsed with MIBK and IPA. Subsequently, a 20 nm thick alumina hard mask is evaporated onto the sample and lifted off. The pattern is transferred to Si with a pseudo-Bosch etch using a mixture of SF$_6$ and C$_4$F$_8$ gases. Finally, the hard mask is removed by immersing the sample in a 1:1 mixture of NH$_4$OH and H$_2$O$_2$.

All simulations, including topology optimization and calculations of resonant modes and transmittance, were performed on a high performance computing cluster. At each iteration, the transmission for every combination of wavelength, angle, and polarization is simulated in parallel with eight Intel CPU cores for each combination (Skylake, Cascadelake, and Icelake processors).

\section{Acknowledgments}
This work was supported by Meta Inc., Army Research Office grant W911NF-22-1-0097, the Caltech Sensing to Intelligence program, AFOSR grant no. FA9550-22-1-0393, and the Simons Collaboration on Extreme Wave Phenomena Based on Symmetries (award no. SFI-MPS-EWP-00008530-09). Phillippe Pearson acknowledges support from the National Sciences and Engineering Research Council of Canada (NSERC). The calculations in this work were performed on a high performance computing cluster as part of the Resnick High Performance Center, which is supported by the Resnick Sustainability Institute.

\textbf{Data availability.} Data underlying the results presented in this paper are not publicly available at this time, but may be obtained from the authors upon reasonable request.

\textbf{Disclosures.} The authors declare no conflicts of interest.

\bibliography{references}

@article{shen_optical_2014,
  author =        {Shen, Yichen and Ye, Dexin and Celanovic, Ivan and
                   Johnson, Steven G. and Joannopoulos, John D. and
                   Soljačić, Marin},
  journal =       {Science},
  month =         mar,
  note =          {Publisher: American Association for the Advancement
                   of Science},
  number =        {6178},
  pages =         {1499--1501},
  title =         {Optical {Broadband} {Angular} {Selectivity}},
  volume =        {343},
  year =          {2014},
  doi =           {10.1126/science.1249799},
  url =           {https://doi.org/10.1126/science.1249799},
}

@article{g_w_mbise_angular_1997,
  author =        {{G W Mbise} and {D Le Bellac} and {G A Niklasson} and
                   {C G Granqvist}},
  journal =       {Journal of Physics D: Applied Physics},
  month =         aug,
  number =        {15},
  pages =         {2103},
  title =         {Angular selective window coatings: theory and
                   experiments},
  volume =        {30},
  year =          {1997},
  doi =           {10.1088/0022-3727/30/15/001},
  issn =          {0022-3727},
  url =           {https://dx.doi.org/10.1088/0022-3727/30/15/001},
}

@article{wesemann_selective_2019,
  author =        {Wesemann, Lukas and Panchenko, Evgeniy and
                   Singh, Kalpana and Della Gaspera, Enrico and
                   Gómez, Daniel E. and Davis, Timothy J. and
                   Roberts, Ann},
  journal =       {APL Photonics},
  month =         oct,
  number =        {10},
  pages =         {100801},
  title =         {Selective near-perfect absorbing mirror as a spatial
                   frequency filter for optical image processing},
  volume =        {4},
  year =          {2019},
  doi =           {10.1063/1.5113650},
  issn =          {2378-0967},
  url =           {https://pubs.aip.org/app/article/4/10/100801/594651/
                  Selective-near-perfect-absorbing-mirror-as-a},
}

@article{cotrufo_dispersion_2023,
  author =        {Cotrufo, Michele and Arora, Akshaj and Singh, Sahitya and
                   Alù, Andrea},
  journal =       {Nature Communications},
  month =         nov,
  number =        {1},
  pages =         {7078},
  title =         {Dispersion engineered metasurfaces for broadband,
                   high-{NA}, high-efficiency, dual-polarization analog
                   image processing},
  volume =        {14},
  year =          {2023},
  doi =           {10.1038/s41467-023-42921-z},
  issn =          {2041-1723},
  url =           {https://www.nature.com/articles/s41467-023-42921-z},
}

@article{qu_polarization-independent_2018,
  author =        {Qu, Yurui and Shen, Yichen and Yin, Kezhen and
                   Yang, Yuanqing and Li, Qiang and Qiu, Min and
                   Soljačić, Marin},
  journal =       {ACS Photonics},
  month =         oct,
  number =        {10},
  pages =         {4125--4131},
  title =         {Polarization-{Independent} {Optical} {Broadband}
                   {Angular} {Selectivity}},
  volume =        {5},
  year =          {2018},
  doi =           {10.1021/acsphotonics.8b00862},
  issn =          {2330-4022, 2330-4022},
  url =           {https://pubs.acs.org/doi/10.1021/acsphotonics.8b00862},
}

@article{xu_broadband_2021,
  author =        {Xu, Jin and Mandal, Jyotirmoy and Raman, Aaswath P.},
  journal =       {Science},
  month =         apr,
  note =          {Publisher: American Association for the Advancement
                   of Science},
  number =        {6540},
  pages =         {393--397},
  title =         {Broadband directional control of thermal emission},
  volume =        {372},
  year =          {2021},
  doi =           {10.1126/science.abc5381},
  url =           {https://doi.org/10.1126/science.abc5381},
}

@article{kwon_nonlocal_2018,
  author =        {Kwon, Hoyeong and Sounas, Dimitrios and
                   Cordaro, Andrea and Polman, Albert and Alù, Andrea},
  journal =       {Physical Review Letters},
  month =         oct,
  number =        {17},
  pages =         {173004},
  title =         {Nonlocal {Metasurfaces} for {Optical} {Signal}
                   {Processing}},
  volume =        {121},
  year =          {2018},
  doi =           {10.1103/PhysRevLett.121.173004},
  issn =          {0031-9007, 1079-7114},
  url =           {https://link.aps.org/doi/10.1103/PhysRevLett.121.173004},
}

@article{kogelnik_coupled_1969,
  author =        {Kogelnik, Herwig},
  journal =       {Bell System Technical Journal},
  month =         nov,
  note =          {Publisher: John Wiley \& Sons, Ltd},
  number =        {9},
  pages =         {2909--2947},
  title =         {Coupled {Wave} {Theory} for {Thick} {Hologram}
                   {Gratings}},
  volume =        {48},
  year =          {1969},
  doi =           {10.1002/j.1538-7305.1969.tb01198.x},
  issn =          {0005-8580},
  url =           {https://doi.org/10.1002/j.1538-7305.1969.tb01198.x},
}

@article{huo_angular_2018,
  author =        {Huo, Pengcheng and Liang, Yuzhang and Zhang, Si and
                   Lu, Yanqing and Xu, Ting},
  journal =       {Laser \& Photonics Reviews},
  month =         aug,
  note =          {Publisher: John Wiley \& Sons, Ltd},
  number =        {8},
  pages =         {1700309},
  title =         {Angular {Optical} {Transparency} {Induced} by
                   {Photonic} {Topological} {Transitions} in
                   {Metamaterials}},
  volume =        {12},
  year =          {2018},
  doi =           {10.1002/lpor.201700309},
  issn =          {1863-8880},
  url =           {https://doi.org/10.1002/lpor.201700309},
}

@article{hamam_angular_2011,
  author =        {Hamam, Rafif E. and Celanovic, Ivan and
                   Soljačić, Marin},
  journal =       {Physical Review A},
  month =         mar,
  number =        {3},
  pages =         {035806},
  title =         {Angular photonic band gap},
  volume =        {83},
  year =          {2011},
  doi =           {10.1103/PhysRevA.83.035806},
  issn =          {1050-2947, 1094-1622},
  url =           {https://link.aps.org/doi/10.1103/PhysRevA.83.035806},
}

@article{xiong_augmented_2021,
  author =        {Xiong, Jianghao and Hsiang, En-Lin and He, Ziqian and
                   Zhan, Tao and Wu, Shin-Tson},
  journal =       {Light: Science \& Applications},
  month =         oct,
  number =        {1},
  pages =         {216},
  title =         {Augmented reality and virtual reality displays:
                   emerging technologies and future perspectives},
  volume =        {10},
  year =          {2021},
  doi =           {10.1038/s41377-021-00658-8},
  issn =          {2047-7538},
  url =           {https://www.nature.com/articles/s41377-021-00658-8},
}

@article{jacob_design_2000,
  author =        {Jacob, Donald K. and Dunn, Steven C. and
                   Moharam, M. G.},
  journal =       {Journal of the Optical Society of America A},
  month =         jul,
  note =          {Publisher: Optica Publishing Group},
  number =        {7},
  pages =         {1241--1249},
  title =         {Design considerations for narrow-band dielectric
                   resonant grating reflection filters of finite length},
  volume =        {17},
  year =          {2000},
  doi =           {10.1364/JOSAA.17.001241},
  url =           {https://opg.optica.org/josaa/abstract.cfm?URI=josaa-17-7-
                  1241},
}

@article{magnusson_physical_2008,
  author =        {Magnusson, Robert and Shokooh-Saremi, Mehrdad},
  journal =       {Optics Express},
  month =         mar,
  note =          {Publisher: Optica Publishing Group},
  number =        {5},
  pages =         {3456--3462},
  title =         {Physical basis for wideband resonant reflectors},
  volume =        {16},
  year =          {2008},
  doi =           {10.1364/OE.16.003456},
  url =           {https://opg.optica.org/oe/abstract.cfm?URI=oe-16-5-3456},
}

@article{c_f_r_mateus_broad-band_2004,
  author =        {{C. F. R. Mateus} and {M. C. Y. Huang} and {Lu Chen} and
                   {C. J. Chang-Hasnain} and {Y. Suzuki}},
  journal =       {IEEE Photonics Technology Letters},
  month =         jul,
  number =        {7},
  pages =         {1676--1678},
  title =         {Broad-band mirror (1.12-1.62 $\mu$m) using a
                   subwavelength grating},
  volume =        {16},
  year =          {2004},
  doi =           {10.1109/LPT.2004.828514},
  issn =          {1941-0174},
}

@article{d_rosenblatt_resonant_1997,
  author =        {{D. Rosenblatt} and {A. Sharon} and {A. A. Friesem}},
  journal =       {IEEE Journal of Quantum Electronics},
  month =         nov,
  number =        {11},
  pages =         {2038--2059},
  title =         {Resonant grating waveguide structures},
  volume =        {33},
  year =          {1997},
  doi =           {10.1109/3.641320},
  issn =          {1558-1713},
}

@article{shokooh-saremi_wideband_2008,
  author =        {Shokooh-Saremi, Mehrdad and Magnusson, Robert},
  journal =       {Optics Express},
  month =         oct,
  note =          {Publisher: Optica Publishing Group},
  number =        {22},
  pages =         {18249--18263},
  title =         {Wideband leaky-mode resonance reflectors: {Influence}
                   of grating profile and sublayers},
  volume =        {16},
  year =          {2008},
  doi =           {10.1364/OE.16.018249},
  url =           {https://opg.optica.org/oe/abstract.cfm?URI=oe-16-22-18249},
}

@article{wang_theory_1993,
  author =        {Wang, S. S. and Magnusson, R.},
  journal =       {Applied Optics},
  month =         may,
  note =          {Publisher: Optica Publishing Group},
  number =        {14},
  pages =         {2606--2613},
  title =         {Theory and applications of guided-mode resonance
                   filters},
  volume =        {32},
  year =          {1993},
  doi =           {10.1364/AO.32.002606},
  url =           {https://opg.optica.org/ao/abstract.cfm?URI=ao-32-14-2606},
}

@article{fan_analysis_2002,
  author =        {Fan, Shanhui and Joannopoulos, J. D.},
  journal =       {Physical Review B},
  month =         jun,
  number =        {23},
  pages =         {235112},
  title =         {Analysis of guided resonances in photonic crystal
                   slabs},
  volume =        {65},
  year =          {2002},
  doi =           {10.1103/PhysRevB.65.235112},
  issn =          {0163-1829, 1095-3795},
  url =           {https://link.aps.org/doi/10.1103/PhysRevB.65.235112},
}

@article{ko_properties_2016,
  author =        {Ko, Yeong Hwan and Niraula, Manoj and Lee, Kyu Jin and
                   Magnusson, Robert},
  journal =       {Optics Express},
  month =         mar,
  note =          {Publisher: Optica Publishing Group},
  number =        {5},
  pages =         {4542--4551},
  title =         {Properties of wideband resonant reflectors under
                   fully conical light incidence},
  volume =        {24},
  year =          {2016},
  doi =           {10.1364/OE.24.004542},
  url =           {https://opg.optica.org/oe/abstract.cfm?URI=oe-24-5-4542},
}

@article{boonruang_broadening_2007,
  author =        {Boonruang, Sakoolkan and Greenwell, Andrew and
                   Moharam, M. G.},
  journal =       {Applied Optics},
  month =         nov,
  note =          {Publisher: Optica Publishing Group},
  number =        {33},
  pages =         {7982--7992},
  title =         {Broadening the angular tolerance in two-dimensional
                   grating resonance structures at oblique incidence},
  volume =        {46},
  year =          {2007},
  doi =           {10.1364/AO.46.007982},
  url =           {https://opg.optica.org/ao/abstract.cfm?URI=ao-46-33-7982},
}

@article{wonjoo_suh_temporal_2004,
  author =        {{Wonjoo Suh} and {Zheng Wang} and {Shanhui Fan}},
  journal =       {IEEE Journal of Quantum Electronics},
  month =         oct,
  number =        {10},
  pages =         {1511--1518},
  title =         {Temporal coupled-mode theory and the presence of
                   non-orthogonal modes in lossless multimode cavities},
  volume =        {40},
  year =          {2004},
  doi =           {10.1109/JQE.2004.834773},
  issn =          {0018-9197},
  url =           {http://ieeexplore.ieee.org/document/1337032/},
}

@book{hecht_optics_2002,
  author =        {Hecht, Eugene},
  publisher =     {Fourth edition. Reading, Mass.: Addison-Wesley},
  title =         {Optics},
  year =          {2002},
  url =           {https://search.library.wisc.edu/catalog/999919136202121},
}

@article{heo_tailoring_2019,
  author =        {Heo, Hyungjun and Lee, Sangjun and Kim, Sangin},
  journal =       {Journal of Lightwave Technology},
  month =         sep,
  number =        {17},
  pages =         {4244--4250},
  title =         {Tailoring {Fano} {Resonance} for {Flat}-{Top}
                   {Broadband} {Reflectors} {Based} on {Single}
                   {Guided}-{Mode} {Resonance}},
  volume =        {37},
  year =          {2019},
  doi =           {10.1109/JLT.2019.2922397},
  issn =          {0733-8724, 1558-2213},
  url =           {https://ieeexplore.ieee.org/document/8735859/},
}

@article{oskooi_meep_2010,
  author =        {Oskooi, Ardavan F. and Roundy, David and
                   Ibanescu, Mihai and Bermel, Peter and
                   Joannopoulos, J.D. and Johnson, Steven G.},
  journal =       {Computer Physics Communications},
  month =         mar,
  number =        {3},
  pages =         {687--702},
  title =         {Meep: {A} flexible free-software package for
                   electromagnetic simulations by the {FDTD} method},
  volume =        {181},
  year =          {2010},
  doi =           {10.1016/j.cpc.2009.11.008},
  issn =          {0010-4655},
  url =           {https://www.sciencedirect.com/science/article/pii/
                  S001046550900383X},
}

@book{sakoda_optical_2005,
  address =       {Berlin, Heidelberg},
  edition =       {2nd},
  editor =        {Sakoda, Kazuaki},
  number =        {80},
  publisher =     {Springer Berlin Heidelberg},
  series =        {Springer {Series} in {Optical} {Sciences}},
  title =         {Optical {Properties} of {Photonic} {Crystals}},
  year =          {2005},
  doi =           {10.1007/b138376},
  isbn =          {978-3-540-20682-8 978-3-540-26965-6},
}

@article{jin_inverse_2020,
  author =        {Jin, Weiliang and Li, Wei and Orenstein, Meir and
                   Fan, Shanhui},
  journal =       {ACS Photonics},
  month =         sep,
  number =        {9},
  pages =         {2350--2355},
  title =         {Inverse {Design} of {Lightweight} {Broadband}
                   {Reflector} for {Relativistic} {Lightsail}
                   {Propulsion}},
  volume =        {7},
  year =          {2020},
  doi =           {10.1021/acsphotonics.0c00768},
  issn =          {2330-4022, 2330-4022},
  url =           {https://pubs.acs.org/doi/10.1021/acsphotonics.0c00768},
}

@article{wang_projection_2011,
  author =        {Wang, Fengwen and Lazarov, Boyan Stefanov and
                   Sigmund, Ole},
  journal =       {Structural and Multidisciplinary Optimization},
  month =         jun,
  number =        {6},
  pages =         {767--784},
  title =         {On projection methods, convergence and robust
                   formulations in topology optimization},
  volume =        {43},
  year =          {2011},
  doi =           {10.1007/s00158-010-0602-y},
  issn =          {1615-147X, 1615-1488},
  url =           {https://link.springer.com/10.1007/s00158-010-0602-y},
}

@article{li_inverse_2022,
  author =        {Li, Zhaoyi and Pestourie, Raphaël and Park, Joon-Suh and
                   Huang, Yao-Wei and Johnson, Steven G. and
                   Capasso, Federico},
  journal =       {Nature Communications},
  month =         may,
  number =        {1},
  pages =         {2409},
  title =         {Inverse design enables large-scale high-performance
                   meta-optics reshaping virtual reality},
  volume =        {13},
  year =          {2022},
  doi =           {10.1038/s41467-022-29973-3},
  issn =          {2041-1723},
  url =           {https://doi.org/10.1038/s41467-022-29973-3},
}

@article{wang_design_2022,
  author =        {Wang, Haiwen and Jin, Weiliang and Guo, Cheng and
                   Zhao, Nathan and Rodrigues, Sean P. and Fan, Shanhui},
  journal =       {ACS Photonics},
  month =         apr,
  number =        {4},
  pages =         {1358--1365},
  title =         {Design of {Compact} {Meta}-{Crystal} {Slab} for
                   {General} {Optical} {Convolution}},
  volume =        {9},
  year =          {2022},
  doi =           {10.1021/acsphotonics.1c02005},
  issn =          {2330-4022, 2330-4022},
  url =           {https://pubs.acs.org/doi/10.1021/acsphotonics.1c02005},
}

@article{pearson_inverse-designed_2025,
  author =        {Pearson, Phillippe and Roberts, Gregory and
                   Faraon, Andrei},
  journal =       {Optica},
  month =         jul,
  note =          {Publisher: Optica Publishing Group},
  number =        {7},
  pages =         {1090--1099},
  title =         {Inverse-designed metasurfaces for multifunctional
                   spatial frequency filtering},
  volume =        {12},
  year =          {2025},
  doi =           {10.1364/OPTICA.560985},
  url =           {https://opg.optica.org/optica/abstract.cfm?URI=optica-12-7-
                  1090},
}

@article{molesky_inverse_2018,
  author =        {Molesky, Sean and Lin, Zin and Piggott, Alexander Y. and
                   Jin, Weiliang and Vucković, Jelena and
                   Rodriguez, Alejandro W.},
  journal =       {Nature Photonics},
  month =         nov,
  number =        {11},
  pages =         {659--670},
  title =         {Inverse design in nanophotonics},
  volume =        {12},
  year =          {2018},
  doi =           {10.1038/s41566-018-0246-9},
  issn =          {1749-4885, 1749-4893},
  url =           {https://www.nature.com/articles/s41566-018-0246-9},
}

@article{su_inverse_2018,
  author =        {Su, Logan and Piggott, Alexander Y. and
                   Sapra, Neil V. and Petykiewicz, Jan and
                   Vučković, Jelena},
  journal =       {ACS Photonics},
  month =         feb,
  number =        {2},
  pages =         {301--305},
  title =         {Inverse {Design} and {Demonstration} of a {Compact}
                   on-{Chip} {Narrowband} {Three}-{Channel} {Wavelength}
                   {Demultiplexer}},
  volume =        {5},
  year =          {2018},
  doi =           {10.1021/acsphotonics.7b00987},
  issn =          {2330-4022, 2330-4022},
  url =           {https://pubs.acs.org/doi/10.1021/acsphotonics.7b00987},
}

@article{fan_broadband_2025,
  author =        {Fan, Ziwei and Hwang, Taeseung and Chen, Yixin and
                   Wong, Zi Jing},
  journal =       {ACS Photonics},
  month =         jun,
  number =        {6},
  pages =         {3117--3123},
  title =         {A {Broadband} {Angular}-{Selective} {Midinfrared}
                   {Photodetector}},
  volume =        {12},
  year =          {2025},
  doi =           {10.1021/acsphotonics.5c00394},
  issn =          {2330-4022, 2330-4022},
  url =           {https://pubs.acs.org/doi/10.1021/acsphotonics.5c00394},
}

@article{kosten_highly_2013,
  author =        {Kosten, Emily D and Atwater, Jackson H and
                   Parsons, James and Polman, Albert and
                   Atwater, Harry A},
  journal =       {Light: Science \& Applications},
  month =         jan,
  number =        {1},
  pages =         {e45--e45},
  title =         {Highly efficient {GaAs} solar cells by limiting light
                   emission angle},
  volume =        {2},
  year =          {2013},
  doi =           {10.1038/lsa.2013.1},
  issn =          {2047-7538},
  url =           {https://www.nature.com/articles/lsa20131},
}

@article{ulbrich_enhanced_2010,
  author =        {Ulbrich, Carolin and Peters, Marius and
                   Bläsi, Benedikt and Kirchartz, Thomas and
                   Gerber, Andreas and Rau, Uwe},
  journal =       {Optics Express},
  month =         jun,
  number =        {S2},
  pages =         {A133},
  title =         {Enhanced light trapping in thin-film solar cells by a
                   directionally selective filter},
  volume =        {18},
  year =          {2010},
  doi =           {10.1364/OE.18.00A133},
  issn =          {1094-4087},
  url =           {https://opg.optica.org/oe/abstract.cfm?uri=oe-18-S2-A133},
}

@article{peters_photonic_2009,
  author =        {Peters, Marius and Goldschmidt, Jan Christoph and
                   Kirchartz, Thomas and Bläsi, Benedikt},
  journal =       {Solar Energy Materials and Solar Cells},
  month =         oct,
  number =        {10},
  pages =         {1721--1727},
  title =         {The photonic light trap—{Improved} light trapping
                   in solar cells by angularly selective filters},
  volume =        {93},
  year =          {2009},
  doi =           {10.1016/j.solmat.2009.05.019},
  issn =          {09270248},
  url =           {https://linkinghub.elsevier.com/retrieve/pii/
                  S0927024809002049},
}

@article{li_switchable_2023,
  author =        {Li, Yannanqi and Semmen, John and Yang, Qian and
                   Wu, Shin-Tson},
  journal =       {Journal of the Society for Information Display},
  month =         may,
  note =          {Publisher: John Wiley \& Sons, Ltd},
  number =        {5},
  pages =         {328--335},
  title =         {Switchable polarization volume gratings for augmented
                   reality waveguide displays},
  volume =        {31},
  year =          {2023},
  doi =           {10.1002/jsid.1200},
  issn =          {1071-0922},
  url =           {https://doi.org/10.1002/jsid.1200},
}

@article{ding_waveguide-based_2023,
  author =        {Ding, Yuqiang and Yang, Qian and Li, Yannanqi and
                   Yang, Zhiyong and Wang, Zhengyang and Liang, Haowen and
                   Wu, Shin-Tson},
  journal =       {eLight},
  month =         dec,
  number =        {1},
  pages =         {24},
  title =         {Waveguide-based augmented reality displays:
                   perspectives and challenges},
  volume =        {3},
  year =          {2023},
  doi =           {10.1186/s43593-023-00057-z},
  issn =          {2662-8643},
  url =           {https://elight.springeropen.com/articles/10.1186/s43593-023-
                  00057-z},
}

@article{rolland_waveguide-based_2024,
  author =        {Rolland, Jannick P. and Goodsell, Jeremy},
  journal =       {Light: Science \& Applications},
  month =         jan,
  number =        {1},
  pages =         {22},
  title =         {Waveguide-based augmented reality displays: a
                   highlight},
  volume =        {13},
  year =          {2024},
  doi =           {10.1038/s41377-023-01371-4},
  issn =          {2047-7538},
  url =           {https://www.nature.com/articles/s41377-023-01371-4},
}

@article{siefke_materials_2016,
  author =        {Siefke, Thomas and Kroker, Stefanie and
                   Pfeiffer, Kristin and Puffky, Oliver and
                   Dietrich, Kay and Franta, Daniel and Ohlídal, Ivan and
                   Szeghalmi, Adriana and Kley, Ernst‐Bernhard and
                   Tünnermann, Andreas},
  journal =       {Advanced Optical Materials},
  month =         nov,
  number =        {11},
  pages =         {1780--1786},
  title =         {Materials {Pushing} the {Application} {Limits} of
                   {Wire} {Grid} {Polarizers} further into the {Deep}
                   {Ultraviolet} {Spectral} {Range}},
  volume =        {4},
  year =          {2016},
  doi =           {10.1002/adom.201600250},
  issn =          {2195-1071, 2195-1071},
  url =           {https://advanced.onlinelibrary.wiley.com/doi/10.1002/
                  adom.201600250},
}

@article{bin-alam_ultra-high-q_2021,
  author =        {Bin-Alam, M. Saad and Reshef, Orad and
                   Mamchur, Yaryna and Alam, M. Zahirul and
                   Carlow, Graham and Upham, Jeremy and
                   Sullivan, Brian T. and Ménard, Jean-Michel and
                   Huttunen, Mikko J. and Boyd, Robert W. and
                   Dolgaleva, Ksenia},
  journal =       {Nature Communications},
  month =         feb,
  number =        {1},
  pages =         {974},
  title =         {Ultra-high-{Q} resonances in plasmonic metasurfaces},
  volume =        {12},
  year =          {2021},
  doi =           {10.1038/s41467-021-21196-2},
  issn =          {2041-1723},
  url =           {https://www.nature.com/articles/s41467-021-21196-2},
}

@article{hrabar_ultra-broadband_2013,
  author =        {Hrabar, Silvio and Krois, Igor and Bonic, Ivan and
                   Kiricenko, Aleksandar},
  journal =       {Applied Physics Letters},
  month =         feb,
  number =        {5},
  pages =         {054108},
  title =         {Ultra-broadband simultaneous superluminal phase and
                   group velocities in non-{Foster} epsilon-near-zero
                   metamaterial},
  volume =        {102},
  year =          {2013},
  doi =           {10.1063/1.4790297},
  issn =          {0003-6951, 1077-3118},
  url =           {https://pubs.aip.org/apl/article/102/5/054108/1068503/Ultra-
                  broadband-simultaneous-superluminal-phase},
}

@article{wang_gain-assisted_2000,
  author =        {Wang, L. J. and Kuzmich, A. and Dogariu, A.},
  journal =       {Nature},
  month =         jul,
  number =        {6793},
  pages =         {277--279},
  title =         {Gain-assisted superluminal light propagation},
  volume =        {406},
  year =          {2000},
  doi =           {10.1038/35018520},
  issn =          {0028-0836, 1476-4687},
  url =           {https://www.nature.com/articles/35018520},
}

@article{lalau-keraly_adjoint_2013,
    title = {Adjoint shape optimization applied to electromagnetic design},
    volume = {21},
    url = {https://opg.optica.org/oe/abstract.cfm?URI=oe-21-18-21693},
    doi = {10.1364/OE.21.021693},
    number = {18},
    journal = {Optics Express},
    author = {Lalau-Keraly, Christopher M. and Bhargava, Samarth and Miller, Owen D. and Yablonovitch, Eli},
    month = sep,
    year = {2013},
    note = {Publisher: Optica Publishing Group},
    pages = {21693--21701},
}

\end{document}